\documentclass[journal]{IEEEtran}
\usepackage{graphicx}
\usepackage{amssymb}
\usepackage{lineno}
\usepackage{hyperref}
\usepackage{amsmath}
\usepackage{algorithm} 
\usepackage{algpseudocode} 
\usepackage{enumerate}
\usepackage{booktabs}
\usepackage{multirow} 
\usepackage{rotating}
\usepackage{pifont}
\usepackage{booktabs} 
\usepackage{tabularx} 
\usepackage{siunitx}  
\usepackage{ragged2e} 
\usepackage{array}    
\usepackage{subcaption}
\usepackage{tabularx} %
\usepackage{caption}  
\usepackage[table,xcdraw]{xcolor}
\usepackage{float}        
\usepackage{subcaption}   
\usepackage{placeins}     
\usepackage{enumitem}
\usepackage{orcidlink}
\usepackage{authblk}
\usepackage{graphicx}
\usepackage{subcaption}  

\newcolumntype{C}{>{\centering\arraybackslash}X}

\newcommand{\xmark}{\ding{55}}%

\newcommand*{\updatedText}[1]{\textcolor{black}{ #1}}
\begin{document}

\title{Deeply Supervised Multi-Task Autoencoder for Biological Brain Age estimation using three dimensional T$_1$-weighted magnetic resonance imaging}

\author[1]{Mehreen~Kanwal}
\author[1]{Yunsik~Son}

\affil[1]{Department of Computer Science and Artificial Intelligence, Dongguk University, Seoul, 04620, Republic of Korea}
\affil[*]{Corresponding E-mail: \href{mailto:sonbug@dgu.ac.kr}{sonbug@dgu.ac.kr}}



\maketitle
\begin{abstract}
Accurate estimation of biological brain age from three dimensional (3D) T$_1$-weighted magnetic resonance imaging (MRI) is a critical imaging biomarker for identifying accelerated aging associated with neurodegenerative diseases. Effective brain age prediction necessitates training 3D models to leverage comprehensive insights from volumetric MRI scans, thereby fully capturing spatial anatomical context. However, optimizing deep 3D models remains challenging due to problems such as vanishing gradients. Furthermore, brain structural patterns differ significantly between sexes, which impacts aging trajectories and vulnerability to neurodegenerative diseases, thereby making sex classification crucial for enhancing the accuracy and generalizability of predictive models. To address these challenges, we propose a Deeply Supervised Multitask Autoencoder (DSMT-AE) framework for brain age estimation. DSMT-AE employs deep supervision, which involves applying supervisory signals at intermediate layers during training, to stabilize model optimization, and multitask learning to enhance feature representation. Specifically, our framework simultaneously optimizes brain age prediction alongside auxiliary tasks of sex classification and image reconstruction, thus effectively capturing anatomical and demographic variability to improve prediction accuracy. We extensively evaluate DSMT-AE on the Open Brain Health Benchmark (OpenBHB) dataset, the largest multisite neuroimaging cohort combining ten publicly available datasets. The results demonstrate that DSMT-AE achieves state-of-the-art performance and robustness across age and sex subgroups. Additionally, our ablation study confirms that each proposed component substantially contributes to the improved predictive accuracy and robustness of the overall architecture.
\end{abstract}

\begin{IEEEkeywords}
Brain age estimation, Multitask learning, Deep Supervision, Magnetic resonance imaging
\end{IEEEkeywords}


\IEEEpeerreviewmaketitle

\section{Introduction}

Accurate prediction of the biological \emph{brain age} of an individual using structural MRI has emerged as a powerful biomarker of brain health and aging \cite{bib_1}. By comparing the predicted brain age to a subject’s chronological age, one obtains a \emph{brain age gap} that quantifies accelerated or decelerated brain aging. A larger positive brain age gap has been linked to accelerated aging, cognitive impairment, and increased risk of neurodegenerative and psychiatric disorders \cite{bib_2,bib_3,bib_4}. These findings underscore the clinical relevance of brain age as an imaging-derived biomarker for distinguishing pathological versus healthy aging. 

Estimation of biological age can be approached using both non-imaging biomarkers and imaging-based techniques. Traditional non-imaging biomarkers include molecular and physiological measures of aging such as epigenetic clocks (DNA methylation patterns) \cite{bib_5}, telomere length \cite{bib_6}, and various blood-based markers. These approaches capture systemic or peripheral aspects of aging but do not directly measure brain structure. In contrast, neuroimaging methods directly assess brain morphology and tissue integrity. High-resolution structural MRI (e.g., T$_1$-weighted scans) provides detailed 3D images of anatomy, enabling quantification of age-related atrophy in cortical and subcortical regions. For example, early work using the BrainAGE framework extracted T$_1$ sMRI features and applied support vector regression to predict chronological age \cite{bib_1}. Other studies have computed voxel-wise or region-of-interest measures (gray matter density, cortical thickness, etc.) and applied regression models (e.g., Gaussian process or SVR) to estimate brain age \cite{bib_7,bib_8}. In this way, imaging-based approaches specialize in capturing brain-specific aging markers that systemic biomarkers may miss. 

Early brain age prediction models relied on handcrafted features and classical machine learning. These models extract summary statistics (e.g., volumes of segmented brain regions or cortical thickness measures from tools like FreeSurfer) and learn a mapping to age using regressors such as support vector machines or random forests \cite{bib_9}. While such handcrafted-feature models provided proof-of-concept evidence that sMRI can predict age, their accuracy is limited by the expressiveness of the chosen features. Consequently, recent approaches have leveraged deep convolutional neural networks (CNNs) to automatically learn hierarchical features directly from raw 3D images \cite{bib_10}. Deep CNN models can capture complex, distributed patterns of brain aging that handcrafted features might miss, often achieving higher prediction accuracy than traditional methods. 

However, training deep 3D CNNs on volumetric MRI data introduces significant challenges. The 3D nature of the input means that networks must process large volumetric inputs, leading to very high-dimensional models with millions of parameters. This increases the risk of overfitting and demands extensive computational and memory resources. Moreover, deep architectures can suffer from vanishing gradients and other optimization difficulties during training if not carefully managed \cite{bib_11,bib_12}. In practice, strategies such as data augmentation, transfer learning, multi-task training, and deep supervision are often employed to mitigate these issues and improve generalization \cite{bib_11,bib_12}. 

\begin{figure*}[ht]
\centering
\includegraphics[width=1\textwidth]{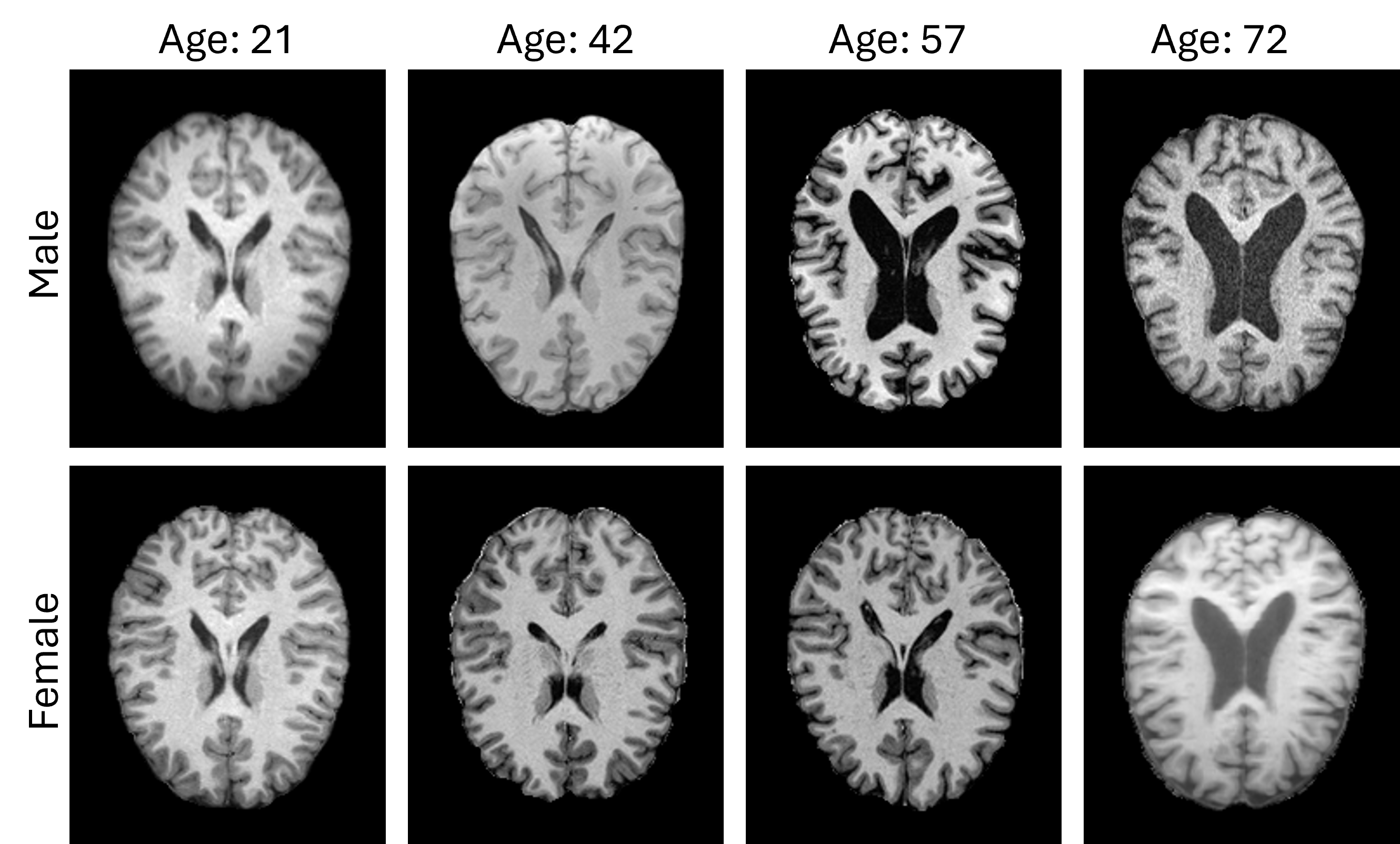}
\caption{Visualization of structural magnetic resonance imaging (sMRI) scans from male and female participants across representative age groups. Each column corresponds to a different age group, highlighting the anatomical variability associated with aging. Notably, structural differences between male and female brains are also evident, underscoring the importance of incorporating sex information in brain age estimation models.}
\label{sex_age_slices}
\end{figure*}

Another important consideration in brain age modeling is the incorporation of demographic variables, particularly biological sex. Male and female brains exhibit well-known anatomical and developmental differences, and their aging trajectories can diverge \cite{bib_13}, as illustrated in Figure~\ref{sex_age_slices}. Ignoring sex-related variation may introduce bias or reduce the accuracy of age predictions. Recent studies have shown that explicitly incorporating sex information can improve brain age estimation performance. For example, some works augment the input data or model features with sex as an additional covariate \cite{bib_14}, while others use a multi-task learning framework to jointly predict age and sex \cite{bib_15}. These multi-task approaches capture sex-specific anatomical variation and have been found to yield more robust and interpretable age predictions. 

Motivated by these observations, we propose a Deep Supervised Multi-Task Autoencoder (DSMT-AE) for brain age estimation from sMRI. Our model is a 3D convolutional autoencoder that compresses a volumetric T$_1$-weighted scan into a low-dimensional latent representation and then reconstructs the input. Crucially, DSMT-AE augments the autoencoder with two supervised output branches: one for age regression and one for sex classification. By jointly predicting age and sex, the network learns shared latent features that capture both general aging patterns and sex-specific anatomical differences. We also incorporate deep supervision by inserting auxiliary age and sex prediction losses at multiple intermediate layers of the network. This encourages the propagation of meaningful gradients to earlier layers and promotes discriminative feature learning at different depths \cite{bib_16}.

In summary, the contributions of this work are as follows:
\begin{enumerate}[label=(\arabic*)]
\item We develop a novel 3D CNN-based autoencoder architecture for brain age prediction from T$_1$-weighted MRI that explicitly incorporates biological sex as a joint prediction task to improve robustness and interpretability.
\item We introduce deep supervision into the autoencoder by applying auxiliary age and sex prediction losses at multiple intermediate layers, facilitating stable optimization and enriched hierarchical feature learning.
\item We propose a multitask learning approach that jointly optimizes brain age prediction, sex classification, and MRI reconstruction tasks, effectively capturing both anatomical variability and demographic influences on brain aging.
\item We conduct comprehensive experiments and evaluations on the large-scale OpenBHB neuroimaging dataset\cite{dufumier2022openbhb}, which integrates ten publicly available multisite datasets, ensuring generalizability and robustness of our findings.
\item We perform extensive ablation studies demonstrating that each proposed component—deep supervision, multitask learning, and the integration of sex classification—significantly contributes to improved predictive accuracy and model robustness across age and sex subgroups.
\end{enumerate}

\section{Related Work}
\label{rw}

Brain age estimation using structural MRI (sMRI) has gained substantial interest as a non-invasive biomarker for assessing brain health, cognitive decline, and neurological disorders. The advent of deep learning has significantly advanced this field by enabling automated extraction and modeling of complex neuroanatomical features directly from imaging data \cite{usman2024advancing, rehman2024biological}. Various architectures have been explored, ranging from convolutional neural networks (CNNs) and attention-based frameworks to graph neural networks (GNNs) and autoencoder-based models.

Initial brain age studies typically employed traditional machine learning methods with handcrafted features extracted from neuroimaging data. For instance, voxel-based morphometry and cortical thickness measures have been widely used alongside support vector regression (SVR) or random forest algorithms \cite{franke2010estimating, baecker2021brain}. However, these approaches suffer from inherent limitations such as reliance on domain knowledge for feature selection and susceptibility to feature extraction biases, leading to limited generalizability.

Deep learning techniques, particularly convolutional neural networks, have overcome many of these limitations by automating hierarchical feature extraction. Early CNN-based methods often applied 2D architectures on individual MRI slices, thus overlooking critical 3D contextual information. Recent studies addressed this gap by adopting 3D CNNs, enabling full exploitation of volumetric data \cite{dinsdale2021learning, feng2020estimating}. Notably, Cheng \textit{et al.} \cite{cheng2021brain} introduced a Two-Stage Age Network (TSAN), progressively refining age estimates through hierarchical feature extraction stages. Despite their effectiveness, deep 3D CNNs face optimization challenges, notably the vanishing gradient problem, hindering training efficiency and feature discrimination \cite{peng2021accurate}.

To improve CNN performance, researchers incorporated advanced architectural enhancements such as attention mechanisms. He \textit{et al.} \cite{he2022global} proposed a global-local transformer model employing attention modules to selectively emphasize informative brain regions, significantly improving prediction accuracy. Although such attention mechanisms improved interpretability and predictive performance, they increased computational complexity, presenting challenges for clinical deployment.

Graph neural networks (GNNs) have also gained attention due to their capacity to model complex structural relationships between brain regions. Studies by Pina \textit{et al.} \cite{pina2022structural} and Xu \textit{et al.} \cite{xu2021brain} leveraged graph-based representations derived from structural and diffusion tensor imaging data, respectively, illustrating the potential of GNNs in capturing intricate anatomical connectivity patterns. Nevertheless, these methods require predefined brain parcellations and sophisticated graph construction processes, limiting scalability and ease of use.

In parallel, autoencoder-based architectures emerged as promising candidates for brain age estimation, combining supervised and unsupervised learning paradigms. Autoencoders learn latent representations through unsupervised reconstruction tasks, implicitly regularizing the learned features. For example, Hu \textit{et al.} \cite{hu2020disentangled} and Cai \textit{et al.} \cite{cai2023graph} employed variational autoencoders (VAEs) to model disentangled latent spaces, enhancing model interpretability. Nonetheless, these frameworks rarely integrated multitask learning or demographic information explicitly, limiting their predictive power.

Recognizing the demographic differences in brain aging patterns, recent studies explicitly incorporated sex as an auxiliary input. Armanious \textit{et al.} \cite{armanious2021age} and Cheng \textit{et al.} \cite{cheng2021brain} demonstrated improvements in brain age prediction accuracy by integrating sex information directly into their CNN models. Similarly, multitask learning (MTL) strategies, which jointly optimize age estimation alongside auxiliary tasks such as sex classification, have shown promising outcomes. Usman \textit{et al.} \cite{usman2024advancing} and Rehman \textit{et al.} \cite{rehman2024biological} applied multitask adversarial autoencoders that leveraged sMRI and functional MRI (fMRI) modalities to integrate complementary imaging information, achieving notable accuracy gains. However, these multimodal approaches depend heavily on preprocessing pipelines and computationally expensive fMRI data, thus limiting their practical clinical applicability. Studies utilizing attention-based and transformer networks for improved segmentation, particularly in challenging modalities, have shown promise in advancing model-based approaches \cite{rehman2024biological, ullah2022cascade}.

Moreover, deep supervision, applying intermediate loss signals at multiple layers of a network, has been extensively applied in medical image segmentation to alleviate training difficulties and accelerate convergence. Surprisingly, despite its proven effectiveness, deep supervision remains underexplored in brain age estimation, with existing studies rarely incorporating it alongside multitask frameworks.

To highlight the current landscape, we summarize key studies in brain age estimation in Table \ref{tab:relatedwork}, outlining their methodological aspects including sex-awareness, unsupervised learning, multitask learning, and deep supervision strategies.

\begin{table*}[ht]
\centering
\caption{Comparison of existing studies on brain age estimation.}
\label{tab:relatedwork}
\begin{tabularx}{\textwidth}{lCCCC}
\toprule
\textbf{Study (Year)} & {Sex-Awareness} & {Unsuperv-ised Task} & {Multitask Learning} & {Deep Supervision} \\
\midrule
He \textit{et al.} (2022) \cite{he2022deep} & \xmark & \xmark & \xmark & \xmark \\
Cheng \textit{et al.} (2021) \cite{cheng2021brain} & \checkmark & \xmark & \xmark & \xmark \\ 
Armanious \textit{et al.} (2021) \cite{armanious2021age} & \checkmark & \xmark & \xmark & \xmark \\ 
Cai \textit{et al.} (2023) \cite{cai2023graph} & \xmark & \checkmark & \xmark & \xmark \\ 
Hu \textit{et al.} (2020) \cite{hu2020disentangled} & \xmark & \checkmark & \xmark &\xmark \\
Liu \textit{et al.} (2023) \cite{liu2023risk} & \xmark & \checkmark & \xmark & \xmark \\
Cheshmi \textit{et al.} (2023) \cite{cheshmi2023brain} & \xmark & \xmark & \xmark &\xmark \\
Adil \textit{et al.} (2024) \cite{aqil2023confounding} & \xmark & \checkmark & \xmark &\xmark \\
Usman \textit{et al.} (2024) \cite{usman2024advancing} & \checkmark & \checkmark & \checkmark &\xmark \\
Rehman \textit{et al.} (2024) \cite{rehman2024biological} & \checkmark & \checkmark & \checkmark &\xmark \\
\midrule
\textbf{Proposed DSMT-AE} & \checkmark & \checkmark & \checkmark & \checkmark \\
\bottomrule
\end{tabularx}
\end{table*}

Given the limitations inherent to multimodal approaches—such as increased complexity, dependency on extensive preprocessing, computational overhead, and limited real-world scalability—there is a clear need for efficient, robust, and scalable single-modality frameworks. To address this, we propose a novel Deeply Supervised Multitask Autoencoder (DSMT-AE) that exclusively utilizes structural MRI data, thereby simplifying the pipeline and enhancing clinical applicability. Our approach combines multitask learning (integrating sex classification and image reconstruction) and deep supervision to explicitly address the challenges of training 3D CNNs on volumetric data, thereby improving model convergence and predictive performance. To the best of our knowledge, this is the first work to simultaneously integrate these strategies within a unified 3D autoencoder framework, significantly advancing the state-of-the-art in brain age estimation from structural MRI alone.

\begin{figure*}[ht]
\centering
\includegraphics[width=1\textwidth]{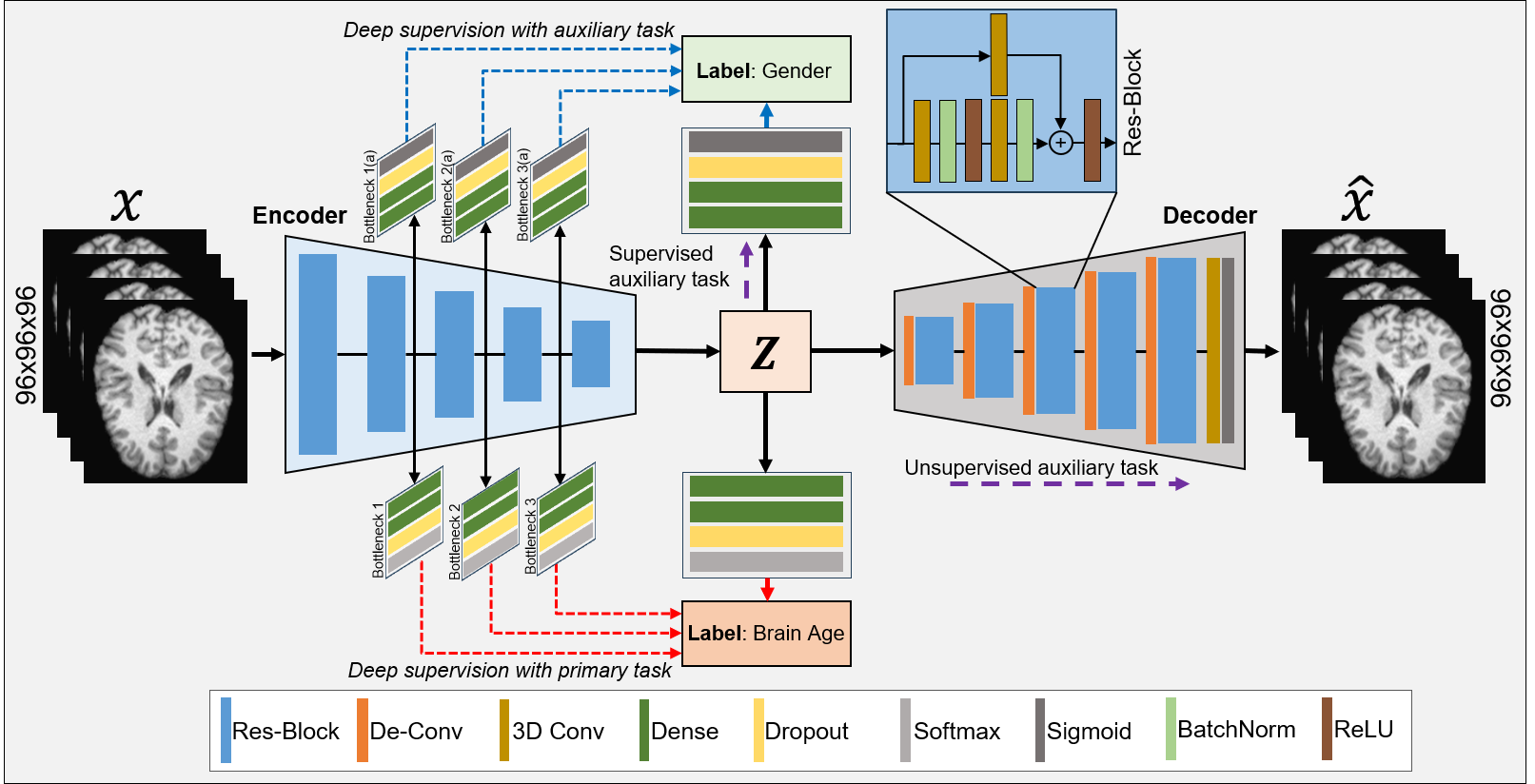}
\caption{Architecture of our proposed Deeply Supervised Multitask Autoencoder (DSMT-AE) for predicting brain age from T1w MRI scans. }
\label{proposed_architecture}
\end{figure*}

\section{Proposed Methodology}
\label{pro}

\subsection{Overall Framework}
In this section, we present the details of our proposed framework for brain age estimation, a 3D deep supervision-based multi-task learning framework. This deeply supervised multi-task autoencoder (DSMT-AE) combines multitask learning with deep supervision, enabling the integration of both supervised and unsupervised auxiliary tasks. Specifically, we introduce gender classification as a supervised auxiliary task, alongside the primary task of brain age estimation, and image reconstruction as an unsupervised auxiliary task. By leveraging these tasks, we enhance the performance of brain age prediction. Below, we elaborate on the architecture and loss formulations.

\subsection{Architectural Details}
The architecture of the proposed framework is illustrated in Figure \ref{proposed_architecture}. Our network consists of an encoder ($E$) and decoder ($D$) with deep supervision applied to intermediate layers through shallow bottleneck branches. The input is a 3D MRI volume of dimensions $96 \times 96 \times 96$. The encoder comprises five 3D residual blocks ($\text{Res-Block}$), where each block has a 3D convolutional layer, followed by batch normalization and an ELU activation function. Deep supervision is applied at three levels of the encoder through bottleneck networks, which perform gender classification and brain age estimation tasks.

The latent representation $z$ produced by the deepest layer of the encoder is fed into both the decoder for the reconstruction task and the auxiliary branches for brain age estimation and gender classification. The decoder consists of five 3D deconvolutional layers, each corresponding to a residual block, and reconstructs the input image $\hat{x}$. The reconstruction task is unsupervised, while the gender classification and brain age estimation tasks are supervised.

Each bottleneck network used for brain age estimation includes two dense layers, a dropout layer, and a final softmax layer for the regression output. For gender classification, the final layer is a sigmoid function. The network is optimized jointly across all tasks with deep supervision to avoid vanishing gradients and enhance feature extraction across the network.

\subsection{Multitask Learning}
In the proposed DSMT-AE, multitask learning is facilitated through the shared encoder $E$ that extracts common features for brain age estimation, gender classification, and image reconstruction. The latent representation $z$ is used for brain age regression and gender classification via auxiliary networks, while it is simultaneously used in the decoder for reconstruction. The total loss function for multitask learning is given by:

\begin{equation}
\label{totalloss}
\mathcal{L}_{\text{DSMT-AE}} = \alpha \mathcal{L}_{\text{AE}} + (1-\alpha) \mathcal{L}_{\text{DST}},
\end{equation}

where $\mathcal{L}_{\text{AE}}$ is the reconstruction loss of the autoencoder and $\mathcal{L}_{\text{DST}}$ represents the combined deep supervision loss for brain age estimation and gender classification. Specifically, $\mathcal{L}_{\text{DST}}$ is defined as:

\begin{equation}
\label{lossc}
\mathcal{L}_{\text{DST}} = \beta \mathcal{L}_{\text{BA}} + (1 - \beta) \mathcal{L}_{\text{GC}},
\end{equation}

where $\mathcal{L}_{\text{BA}}$ is the brain age loss, and $\mathcal{L}_{\text{GC}}$ is the gender classification loss. The trade-off parameters $\alpha$ and $\beta$ control the contribution of each task to the total loss.

\subsection{Deep Supervision}
Deep supervision is applied at multiple stages of the encoder to avoid gradient vanishing and accelerate convergence. For both brain age estimation and gender classification, we introduce shallow bottleneck branches at intermediate layers of the encoder. The brain age loss and gender classification loss are formulated as:

\begin{equation}
\label{loss_ba}
\mathcal{L}_{\text{BA}} = \gamma \mathcal{L}_{\text{BA}_f} + (1 - \gamma) \sum_{d \in \mathcal{D}} \eta_d \mathcal{L}_{\text{BA}_d},
\end{equation}

\begin{equation}
\label{loss_gc}
\mathcal{L}_{\text{GC}} = \gamma \mathcal{L}_{\text{GC}_f} + (1 - \gamma) \sum_{d \in \mathcal{D}} \eta_d \mathcal{L}_{\text{GC}_d},
\end{equation}

where $\mathcal{L}_{\text{BA}_f}$ and $\mathcal{L}_{\text{GC}_f}$ are the losses from the final, deepest, bottleneck for brain age and gender classification, respectively, and $\mathcal{L}_{\text{BA}_d}$ and $\mathcal{L}_{\text{GC}_d}$ are the losses from the $d$-th intermediate shallow bottleneck. The parameters $\gamma$ and $\eta_d$ control the contributions of the final and intermediate layers.

\subsection{Self-Ensemble Mechanism}
Ensembling is a commonly used technique in deep learning that enhances performance by combining the outputs of multiple models or averaging the weights of independently trained models, thereby improving model robustness and generalisation across diverse real-world scenarios \cite{Breiman2001, Schamoni2022EnsemblingNN}. However, ensembled deep learning models often require substantial computational resources, which makes their deployment in real-time clinical settings challenging. 

To address this, a self-ensembling scheme has been proposed in the literature \cite{french2018selfensembling, PERONE20191}, which avoids the computational overhead by using different iterations or versions of the same model framework to combine multiple predictions. In this work, we employ self-ensembling during inference by weighted averaging of outputs obtained from both the main and shallow regressors in our proposed framework. This approach enhances the model's predictive capability without significantly increasing computational cost. 

Our self-ensemble strategy is defined as follows:

\begin{equation}
\label{loss_gc}
\hat{y}_{\text{BA}_{ensemble}} = \rho \hat{y}_{\text{BA}_f} + (1 - \rho) \sum_{d \in \mathcal{D}} \Omega_d ~\hat{y}_{\text{BA}_d},
\end{equation}

where $\hat{y}_{\text{BA}_f}$ represents the predicted brain age from the deepest, final regressor, and $\hat{y}_{\text{BA}_d}$ corresponds to the predictions from the $d^{th}$ shallow regressor. The parameters $\rho$ and $\Omega_d$ are weights empirically determined using the validation set.

\subsection{Loss Functions}
The total loss function in our framework is a weighted sum of three components: the reconstruction loss, brain age estimation loss, and gender classification loss. This joint loss optimizes the encoder ($E$), decoder ($D$), and auxiliary branches. It can be represented as:

\begin{equation}
\label{totalloss_extended}
\mathcal{L}_{\text{total}} = \alpha \mathcal{L}_{\text{AE}} + (1-\alpha)(\beta \mathcal{L}_{\text{BA}} + (1 - \beta) \mathcal{L}_{\text{GC}}),
\end{equation}

where $\mathcal{L}_{\text{AE}}$ is the reconstruction loss, $\mathcal{L}_{\text{BA}}$ is the brain age estimation loss (MAE), and $\mathcal{L}_{\text{GC}}$ is the binary cross-entropy loss for gender classification. The weights $\alpha$ and $\beta$ control the contribution of each loss to the total objective.

The reconstruction loss ensures that the decoder ($D$) accurately reconstructs the input image from the latent representation $z$. It is computed as the mean squared error (MSE) between the input image $x$ and the reconstructed output $\hat{x}$:

\begin{equation}
\label{AE_loss}
\mathcal{L}_{\text{AE}}(x, \hat{x}) = \lVert x - \hat{x} \rVert_2^2,
\end{equation}

where $\hat{x} = D(E(x))$.

For brain age estimation, we use the mean absolute error (MAE) to penalize the absolute difference between the predicted brain age $\hat{y}_{\text{BA}}$ and the chronological brain age $y_{\text{BA}}$:

\begin{equation}
\label{MAE_loss}
\mathcal{L}_{\text{BA}}(y_{\text{BA}}, \hat{y}_{\text{BA}}) = \lVert y_{\text{BA}} - \hat{y}_{\text{BA}} \rVert_1.
\end{equation}

The binary cross-entropy loss is used for the gender classification task, which measures the difference between the predicted probability $\hat{p}_{\text{GC}}$ and the true label $y_{\text{GC}}$:

\begin{equation}
\label{BCE_loss}
\mathcal{L}_{\text{GC}}(y_{\text{GC}}, \hat{p}_{\text{GC}}) = - \left[ y_{\text{GC}} \log(\hat{p}_{\text{GC}}) + (1 - y_{\text{GC}}) \log(1 - \hat{p}_{\text{GC}}) \right].
\end{equation}

The final combined loss for training the model is:

\begin{multline}
\mathcal{L}_{\text{total}} = \alpha \lVert x - \hat{x} \rVert_2^2 
+ \beta \lVert y_{\text{BA}} - \hat{y}_{\text{BA}} \rVert_1 \\
+ (1 - \beta) \left[ y_{\text{GC}} \log(\hat{p}_{\text{GC}}) + (1 - y_{\text{GC}}) \log(1 - \hat{p}_{\text{GC}}) \right]
\end{multline}

where $\alpha$ and $\beta$ balance the contributions of the reconstruction, brain age estimation, and gender classification losses.

\begin{figure*}[ht]
\centering
\includegraphics[width=1\textwidth]{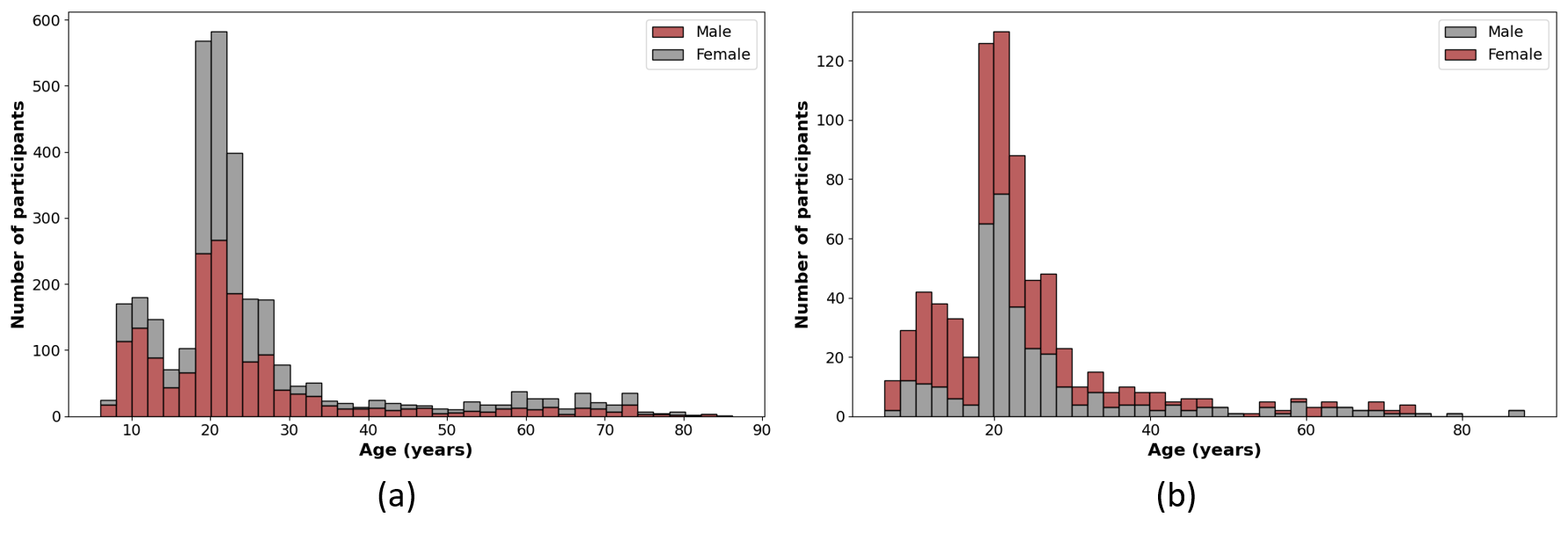}
\caption{\updatedText{Age distribution of male and female participants in the training and validation sets of the OpenBHB dataset\cite{dufumier2022openbhb} , shown in (a) and (b), respectively.}}
\label{data_distribution}
\end{figure*}


\section{Experimental Details}
\subsection{Dataset and Pre-processing}
\label{dataset_section}
This study utilises the Open Big Healthy Brains (OpenBHB) dataset \cite{dufumier2022openbhb}, which includes over 5,000 3D T1-weighted (T1w) brain MRIs from healthy controls (HC). The primary challenge in this context is modelling normal brain development by building a robust brain age predictor. OpenBHB aggregates data from ten publicly available datasets, extracting only the HC participants from each. Specifically, OpenBHB consists of \( N = 5,330 \) 3D T1w brain MRI scans from HC, acquired from 71 different sites, with some sites using multiple acquisition protocols. 

The OpenBHB dataset promotes diversity, as it includes participants from European-American, European, and Asian genetic backgrounds. To manage redundancy, only one session per participant was retained, along with its best-associated run, selected based on image quality. In addition to the images, the OpenBHB dataset provides associated participant phenotypes as well as site and scanner information, including age, sex, acquisition site, diagnosis, MRI scanner magnetic field strength, and scanner settings.

All data were uniformly pre-processed using the \textit{brainprep}\footnote{https://brainprep.readthedocs.io/en/latest/} module, which employs container technologies such as Quasi-Raw, CAT12\footnote{https://neuro-jena.github.io/cat//} Voxel-Based Morphometry (VBM), and FreeSurfer\footnote{https://surfer.nmr.mgh.harvard.edu/}. A semi-automatic quality control (QC) process was also performed on the dataset, guided by quality metrics. This QC process used the average correlation (via Fisher's z transform) between registered images, retaining only images with a correlation threshold above 0.5.

In this study, we used only the Quasi-Raw scans, which are minimally pre-processed. These scans were generated using ANTs bias field correction, FSL FLIRT\footnote{https://fsl.fmrib.ox.ac.uk/fsl/docs/} 9 degrees of freedom (without shearing) affine registration to the isotropic 1mm MNI template, followed by the application of brain masks.

The OpenBHB dataset is divided into public and private subsets. The public data is further split into training and validation sets, which are useful for cross-validation and for generating publicly comparable results. The private data is reserved for scoring models submitted to the OpenBHB challenge. In this study, we used only the public dataset, which contains 3,966 scans. These were divided into a training set (3,227 scans) and a validation set (757 scans), with a balanced gender distribution across ten age bins (male: 2,080; female: 1,886) (see Figure \ref{data_distribution}).

\subsection{Experimental Settings}
All experiments were conducted on a machine running Ubuntu 20.04.5 LTS, powered by an AMD EPYC 7502 32-Core Processor with 16 available cores, clocked at 2.50 GHz. The system was equipped with 64GB of memory (4 × 16GB DIMM DRAM Synchronous modules) and utilised an NVIDIA A40 GPU with 46GB of VRAM for computation. The model was trained for 200 epochs with a batch size of 4 for training and 2 for validation. We used the Adam optimiser with an initial learning rate of 0.001, which was adjusted using a cosine decay schedule throughout the training process. An early stopping mechanism was employed to halt training if the validation MAE for age prediction did not improve for 20 consecutive epochs, and the best model weights were restored. The weighting coefficients $\alpha$, $\beta$, and $\gamma$ were optimally fine-tuned via a coarse-to-fine grid search on the validation split to balance brain-age regression loss for best brain age estimation performance. To enhance the model's generalisability, data augmentation was applied during training, including random flipping along the x, y, and z axes, random rotation of the 3D volumes within a range of -20 to 20 degrees, zooming with random scaling factors between 0.9 and 1.1, and random erasing where a random cubic region of the volume was set to zero to simulate missing parts of the scan. After augmentation, the 3D volumes were cropped to remove the blank region around the brain and downsampled to a consistent size of 96 × 96 × 96 voxels. Finally, the volumes were normalised to the range [0, 1] before being input to the model. The source code is publicly available at \url{https://github.com/PLASS-Lab/DSMTA}

\subsection{Evaluation Parameters}

In this study, we used three key evaluation metrics to quantitatively assess the performance of our biological brain age estimation model. These metrics include the Mean Absolute Error ($MAE$) with standard deviation (SD), root mean square error ($RMSE$), and the correlation coefficient ($R^2$) between the predicted biological brain ages and the ground-truth chronological brain age labels. Each of these metrics provides insight into different aspects of the model’s performance.


The Mean Absolute Error (MAE) is used to measure the average magnitude of the errors between the predicted brain ages (\(\hat{y}\)) and the ground-truth chronological brain ages (\(y\)). It provides a clear indication of how close the predictions are to the actual values, with lower values indicating better performance. Mathematically, MAE is defined as:

\begin{equation}
\text{MAE} = \frac{1}{N} \sum_{i=1}^{N} \lvert y_i - \hat{y}_i \rvert,
\end{equation}

where \(N\) is the number of samples, \(y_i\) is the true brain age, and \(\hat{y}_i\) is the predicted brain age. MAE penalises all errors equally and is particularly useful in biological brain age estimation, as it provides an intuitive measure of the overall prediction error.

The standard deviation (SD) of the error quantifies the dispersion or variability of the prediction errors. It measures how much the individual prediction errors deviate from the mean error, providing insight into the consistency of the model’s predictions. The SD of the error is defined as:

\begin{equation}
\text{SD} = \sqrt{\frac{1}{N} \sum_{i=1}^{N} \left( (y_i - \hat{y}_i) - \overline{e} \right)^2},
\end{equation}

where \(\overline{e}\) is the mean error:

\begin{equation}
\overline{e} = \frac{1}{N} \sum_{i=1}^{N} (y_i - \hat{y}_i).
\end{equation}

A lower SD indicates that the prediction errors are more concentrated around the mean error, suggesting that the model’s predictions are consistent across different samples.

The Root Mean Square Error ($RMSE$) is a common metric used to measure the accuracy of predictions. It provides a single value that aggregates both the magnitude and variance of the prediction errors. Unlike MAE, RMSE penalises larger errors more heavily, making it sensitive to outliers. The RMSE is defined as:

\begin{equation}
\text{RMSE} = \sqrt{\frac{1}{N} \sum_{i=1}^{N} (y_i - \hat{y}_i)^2}.
\end{equation}

RMSE provides a comprehensive measure of the model’s predictive performance, as it combines both the magnitude and dispersion of the errors into one value. Lower RMSE values indicate better model performance.


Finally, we use Coefficient of Determination (\(R^2\)) which quantifies the proportion of variance in the true ages explained by the predictions.  

\begin{equation}
  R^2
  = 1 \;-\; \frac{\displaystyle\sum_{i=1}^N (y_i - \hat y_i)^2}
                   {\displaystyle\sum_{i=1}^N (y_i - \bar y)^2}\,.
\end{equation}

A value of \(R^2=1\) denotes perfect prediction, \(R^2=0\) indicates that the model performs no better than predicting the sample mean, and negative values arise when the model fits worse than the mean baseline.

Lower MAE and RMSE indicate more accurate predictions, while a higher \(R^2\) (closer to 1) signifies stronger overall agreement between predicted and actual brain ages.

\section{Results and Discussion}
\label{res}

\begin{table*}[ht]
\centering
\caption{Comparison of brain‐age estimation performance (MAE in years) on the OpenBHB dataset.}
\begin{tabular}{l l c}
\toprule
\textbf{Study (Year)} & \textbf{Method}                                    & \textbf{MAE} \\
\midrule
Aqil \textit{et al.} \cite{aqil2023confounding}, 2023  
    & Modified 2D Autoencoder                         & 4.55         \\
Ahmed \textit{et al.} \cite{ahmed2023robust}, 2023
    & Classical regression (handcrafted features)     & 3.25         \\
Cheshmi \textit{et al.} \cite{cheshmi2023brain}, 2023 
    & 3D ResNet‐18 (federated learning)                & 3.86         \\
Träuble \textit{et al.} \cite{trauble2024contrastive}, 2024 
    & 3D ResNet‐18 + contrastive loss                  & 3.72         \\
Usman \textit{et al.} \cite{usman2024advancing}, 2024 
    & Multi‐modal adversarial MTL autoencoder         & 2.77         \\
Rehman \textit{et al.} \cite{rehman2024biological}, 2025 
    & Sex‐aware dual‐path autoencoder                  & 2.72         \\ 
\midrule
\textbf{Ours (DSMT‐AE)} 
    & Deeply supervised multi‐task 3D autoencoder      & \textbf{2.64} \\
\bottomrule
\end{tabular}
\label{tab_SOTA_comparison}
\end{table*}

\subsection{Comparison with State‐of‐the‐Art Methods}

To benchmark our Deeply Supervised Multitask Autoencoder (DSMT‐AE), we compare its performance on the OpenBHB dataset~\cite{dufumier2022openbhb} against recent published approaches that report results on the same data split. Table~\ref{tab_SOTA_comparison} summarizes each method’s architecture and corresponding brain age estimation performance in term of mean absolute error (MAE). Our DSMT‐AE attains an MAE of 2.64 years, surpassing all prior methods.  Compared to classical regression on handcrafted features~\cite{ahmed2023robust}, our end‐to‐end 3D framework learns richer representations directly from volumetric MRI, yielding a 19\% MAE reduction.  Relative to standard and contrastive 3D ResNets~\cite{cheshmi2023brain,trauble2024contrastive}, DSMT‐AE’s integration of auxiliary tasks (sex classification and reconstruction) and deep supervision mitigates gradient vanishing and encourages discriminative feature learning, improving MAE by up to 30\%.  

Against recent multi‐modal adversarial autoencoders~\cite{usman2024advancing,rehman2024biological}, which fuse sMRI and fMRI and incorporate sex information, our model’s deeply supervised multitask design more effectively disentangles sex‐specific aging patterns within a single sMRI stream, yielding a further 3–5\% MAE improvement.  This result demonstrates that combining multi‐task loss weighting with hierarchical supervision can extract more robust, age‐relevant features without the added complexity or computational cost of multiple imaging modalities.  

Although our primary comparison involves state-of-the-art (SOTA) studies utilizing the OpenBHB dataset \cite{dufumier2022openbhb}, numerous prior studies have used different datasets or combinations thereof. To provide a comprehensive overview and contextualize our results, we present a comparative analysis of our proposed method against these studies as summarized in Table~\ref{tab:brain_age}. The OpenBHB dataset was selected due to its integration of ten widely recognized, publicly available neuroimaging datasets, detailed in Section \ref{dataset_section}.

As demonstrated in Table~\ref{tab:brain_age}, our proposed method achieved the lowest mean absolute error (MAE) among the compared studies, with an MAE of 2.64. Notably, our method attained this performance despite utilizing fewer scans (5,330) compared to other high-performing studies \cite{he2021global, yang2022estimating, siegel2025transformers, cai2022graph, dartora2024deep}, which employed significantly larger datasets but exhibited comparable or marginally higher MAEs. In contrast, studies that utilized smaller datasets \cite{zhang2024triamese, gianchandani2023voxel, eltashani2025brain} reported higher MAEs, indicating challenges in performance scalability with limited data. Moreover, several recent studies \cite{he2021global, yang2022estimating, cai2022graph, dartora2024deep, zhang2024triamese} have employed transformer-based models or hybrid architectures incorporating transformers. These architectures, while effective, are known for their substantial computational demands. Siegel et al.~\cite{siegel2025transformers} explicitly compared CNN and transformer-based models, demonstrating that a conventional CNN (ResNet) achieved superior MAE performance (2.66) compared to Swin Transformer (2.67) and Vision Transformer (3.02), emphasizing CNN's computational efficiency and effectiveness.

The key insight derived from these comparative analyses is that carefully designed 3D CNN-based architectures, particularly when combined with multitask learning and deep supervision strategies, can deliver state-of-the-art brain-age estimation performance even with comparatively limited datasets, as demonstrated by our proposed 3D deeply supervised multitask autoencoder.

\begin{table*}[htbp]
\centering
\caption{Comparison of brain‐age estimation performance on various datasets in term of mean absolute error (MAE).}
\small
\scalebox{0.9}{
\begin{tabular}{@{}p{6.5cm}p{5cm}p{3cm}p{1.5cm}@{}}
\toprule
\textbf{Study, year} & \textbf{Dataset} & \textbf{No. of Scans} & \textbf{MAE} \\
\midrule
He et al.\ (2022) \cite{he2021global}            & Multiple Cohort                                                         & 8,379       & 2.70    \\
Yang et al.\ (2022) \cite{yang2022estimating}    & Multiple Cohort                                             & 22,645      & 2.855  \\
Siegel et al.\ (2023) \cite{siegel2025transformers} & UKB \cite{sudlow2015uk}                                                        & 45,209      & 2.67   \\
Cai et al.\ (2023) \cite{cai2022graph}            & UKB \cite{sudlow2015uk}+ ADNI \cite{jack2008alzheimer}                                                 & 16,458      & 2.71   \\
Dartora et al.\ (2024) \cite{dartora2024deep}     & Multiple Cohorts                                              & 17,296      & 2.67   \\
Zhang et al.\ (2024) \cite{zhang2024triamese}     & IXI \cite{ixi_dataset}+ ADNI \cite{jack2008alzheimer}                                                & 1,351 & 3.87   \\
Gianchandani et al.\ (2024) \cite{gianchandani2023voxel} & Cam‑CAN \cite{taylor2017cambridge} + CC359 \cite{souza2018open} & 1,010   & 5.30   \\
Eltashani et al.\ (2025) \cite{eltashani2025brain} & Multiple Cohort                                            & 2,251       & 3.557  \\
Our proposed Method (2025)                         & OpenBHB \cite{dufumier2022openbhb}                         & 5,330       & 2.64   \\
\bottomrule
\end{tabular}
}
\label{tab:brain_age}
\end{table*}

In summary, DSMT‐AE sets a new state of the art on OpenBHB by achieving the lowest MAE while maintaining a single‐modality input and real‐time inference potential, making it well suited for large‐scale clinical deployment.

\begin{table*}[ht]
\centering
\caption{Performance comparison of different models evaluated using Mean Absolute Error (\textit{MAE}), Root Mean Square Error (\textit{RMSE}), and Coefficient of Determination (\(R^2\)) for overall, male, and female scans. \textit{The best results are underlined.}}
\scalebox{0.9}{
\begin{tabular}{@{}lccccccccc@{}}
\toprule
\updatedText{\textbf{Model Name}} & \multicolumn{3}{c}{\updatedText{\textbf{Overall}}} & \multicolumn{3}{c}{\updatedText{\textbf{Male}}} & \multicolumn{3}{c}{\updatedText{\textbf{Female}}} \\ 
\cmidrule(lr){2-4} \cmidrule(lr){5-7} \cmidrule(lr){8-10}
 & \updatedText{\textit{MAE} $\downarrow$} & \updatedText{\textit{RMSE} $\downarrow$} & \updatedText{\(R^2\uparrow\)} & \updatedText{\textit{MAE} $\downarrow$} & \updatedText{\textit{RMSE} $\downarrow$} & \updatedText{\(R^2\uparrow\)} & \updatedText{\textit{MAE} $\downarrow$} & \updatedText{\textit{RMSE} $\downarrow$} & \updatedText{\(R^2\uparrow\)} \\ 
\midrule
\updatedText{Baseline}            & \updatedText{3.93 $\pm$ 4.60} & \updatedText{6.05} & \updatedText{0.77} & \updatedText{3.47 $\pm$ 3.60} & \updatedText{5.00} & \updatedText{0.82} & \updatedText{4.51 $\pm$ 5.54} & \updatedText{7.14} & \updatedText{0.73} \\
\updatedText{AE}           & \updatedText{3.49 $\pm$ 4.12}& \updatedText{5.40} & \updatedText{0.82} & \updatedText{3.20 $\pm$ 3.41} & \updatedText{4.68} & \updatedText{0.84} & \updatedText{3.85 $\pm$ 4.83} & \updatedText{6.17} & \updatedText{0.80} \\
\updatedText{MTL-AE}           & \updatedText{2.97 $\pm$ 1.73} & \updatedText{3.44} & \updatedText{0.93} & \updatedText{2.86 $\pm$ 1.57} & \updatedText{3.27} & \updatedText{0.92} & \updatedText{3.11 $\pm$ 1.89} & \updatedText{3.64} & \updatedText{0.93} \\
\updatedText{DS-AE}          & \updatedText{2.86 $\pm$ 1.66} & \updatedText{3.31} & \updatedText{0.93} & \updatedText{2.82 $\pm$ 1.58} & \updatedText{3.23} & \updatedText{0.92} & \updatedText{2.92 $\pm$ 1.74} & \updatedText{3.40} & \updatedText{0.94} \\
\updatedText{DSMT-AE}       & \updatedText{\underline{2.64 $\pm$ 1.58}} & \updatedText{\underline{3.08}} & \updatedText{\underline{0.94}} & \updatedText{\underline{2.65 $\pm$ 1.57}} & \updatedText{\underline{3.08}} & \updatedText{\underline{0.93}} & \updatedText{\underline{2.64 $\pm$ 1.58}} & \updatedText{\underline{3.07}} & \updatedText{\underline{0.95}} \\
\bottomrule
\end{tabular}
}
\label{ablation_tab}
\end{table*}

\subsection{Ablation Study}

To quantify the individual contributions of reconstruction, multitask learning, and deep supervision in our DSMT‐AE framework, we performed a stepwise ablation study (Table~\ref{ablation_tab}).  Beginning with a 3D ResNet “baseline” trained solely on age regression, we then (1) added an autoencoder branch for unsupervised reconstruction (AE), (2) incorporated a gender‐classification auxiliary task into a multitask AE (MTL‐AE), (3) applied deep supervision to intermediate encoder bottlenecks (DS‐AE), and finally (4) combined both multitask learning and deep supervision in the full DSMT‐AE model. Table~\ref{ablation_tab} presents a controlled sequence of model variants that isolate the effects of unsupervised reconstruction, sex‐aware multitask learning, and deep supervision.  Simply adding the autoencoder branch (AE) to the ResNet backbone yields a noticeable reduction in prediction error and an increase in explained variance, suggesting that enforcing image reconstruction encourages the encoder to learn more anatomically meaningful features.  Introducing the gender‐classification task (MTL‐AE) further regularizes the feature space: by requiring the network to distinguish male and female patterns, it sharpens age‐related representations and markedly improves both accuracy and consistency across subjects.  Applying deep supervision alone (DS‐AE) likewise enhances performance, indicating that intermediate loss signals help mitigate vanishing gradients and promote discriminative feature extraction at multiple depths.

When combined in the full DSMT‐AE architecture, these components produce a synergistic effect: reconstruction constraints provide robust low‐level structure, the sex‐classification auxiliary task regularizes demographic variability, and hierarchical supervision ensures strong gradient flow throughout the encoder.  The result is a model that not only achieves the best overall error statistics but also maintains balanced performance across male and female cohorts.  This ablation confirms that each design choice contributes complementary strengths—together yielding a more reliable and generalizable brain‐age estimator than any single technique alone.

\subsection{Robustness Analysis}

To rigorously assess the robustness of our proposed DSMT-AE model, we compared it against four baseline variants under identical training and evaluation protocols on the OpenBHB test split (10\% hold-out). The baselines included: a vanilla 3D ResNet-18 regression network trained solely on age; a single-task 3D autoencoder (AE) with reconstruction plus a final age regressor; a multitask AE augmented with a gender-classification branch (MT-AE); and a deeply supervised AE incorporating intermediate bottleneck losses but no sex information (DS-AE). For each model, we visualized predicted versus true chronological age, color‐coded by sex, with the identity line (\(\hat y = y\)) overlaid along with the 95\% confidence band (±1.96 × SD of residuals). In each scatter plot we also annotated the mean absolute error (MAE) and its standard deviation (SD), so that both average accuracy and prediction consistency could be directly compared.

The ResNet-18 baseline exhibited the highest MAE (3.93 ± 4.60 yrs) and the widest confidence band, indicating large bias and considerable variance in its estimations. Introducing unsupervised reconstruction via an autoencoder reduced the MAE to 3.49 ± 4.12 yrs, but the confidence band remained broad, suggesting only modest gains in stability. In contrast, augmenting the AE with a gender-classification auxiliary task (MT-AE) sharply lowered both the MAE (2.97 ± 1.73 yrs) and the width of the confidence interval. This improvement demonstrates that sex information serves as a powerful regularizer, guiding the network to learn features that are simultaneously informative of age and sex. Similarly, incorporating deep supervision alone (DS-AE) yielded comparable accuracy (MAE = 2.86 ± 1.66 yrs) and further tightened the confidence band, confirming that direct gradient signals at intermediate encoder layers mitigate vanishing-gradient issues and foster more discriminative latent representations.

Our full DSMT-AE model, which combines both sex-aware multitask learning and deep supervision, achieved the lowest MAE (2.64 ± 1.58 yrs) and the narrowest 95$\%$ confidence interval of all variants. These results highlight two key findings: first, multitask regularization via gender classification significantly enhances both accuracy and consistency; and second, deep supervision accelerates convergence and stabilizes feature learning across the encoder hierarchy. By jointly optimizing for reconstruction, gender classification, and age regression at multiple scales, DSMT-AE constructs a latent embedding that is both highly predictive and robust, making it well-suited for reliable brain‐age estimation in clinical and large‐scale research settings.

\begin{figure*}[t]
\centering
\includegraphics[width=1\textwidth]{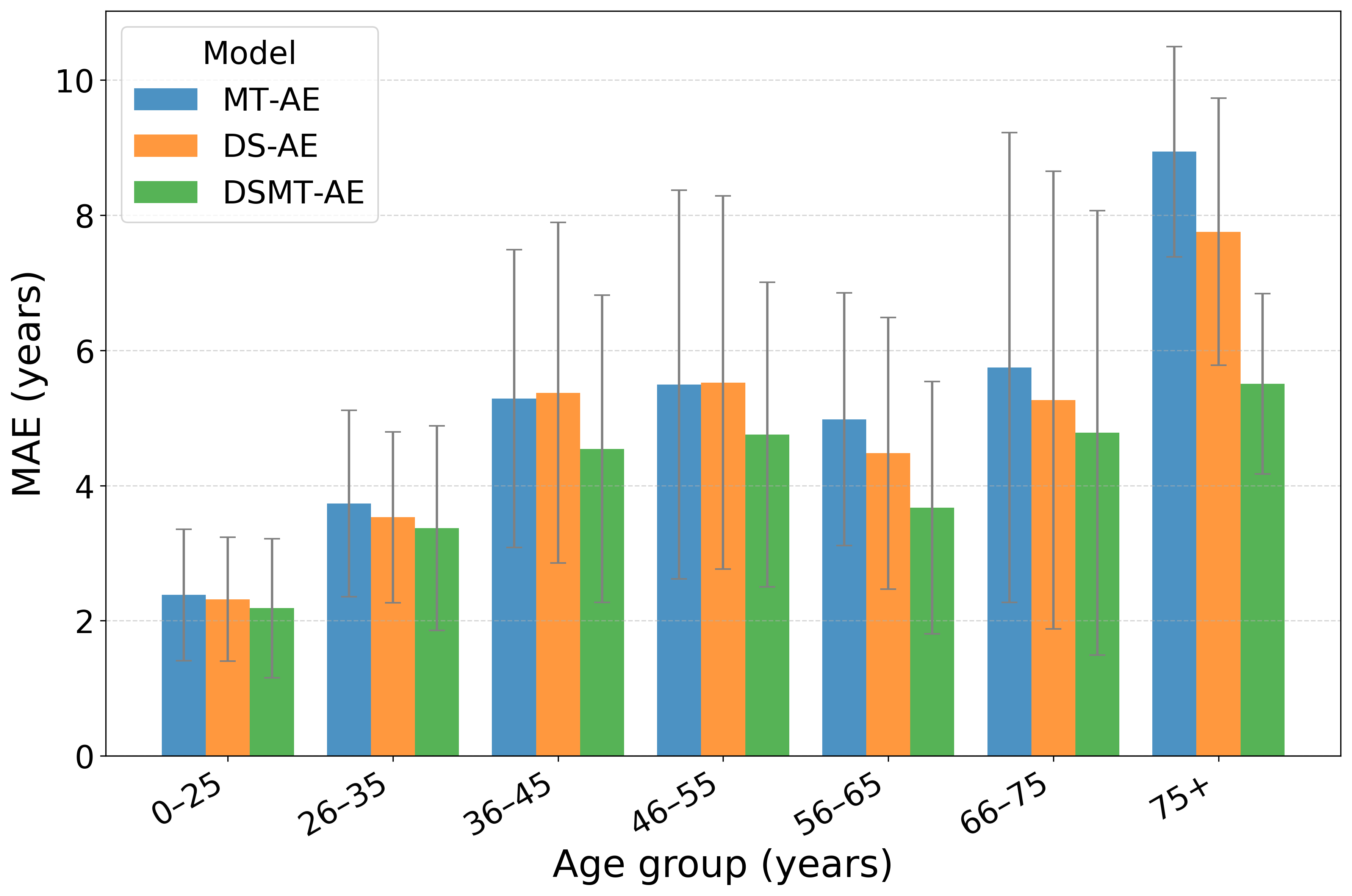}
\caption{Mean Absolute Error (MAE) for brain age estimation obtained from three models, i.e., Multi-task Autoencoder (MT-AE), Deeply Supervised Autoencoder (DS-AE), and our proposed Deeply supervised multitask Autoencoder (DSMT-AE), on difference age groups. }
\label{results_bar_chart}
\end{figure*}

To further demonstrate robustness, Figure~\ref{results_bar_chart} disaggregates prediction error by seven age brackets: \(\le\)25, 26–35, 36–45, 46–55, 56–65, 66–75, and above 75 years.  In each bin we report the mean absolute error (MAE) with its standard‐deviation error bar for three architectures: the basic multitask autoencoder (MT‐AE), the deeply supervised autoencoder without sex conditioning (DS‐AE), and our full deeply supervised multitask autoencoder (DSMT‐AE).  Across all age groups, DSMT‐AE achieves the lowest MAE, with the largest gains seen in older cohorts: for subjects over 75 years, DSMT‐AE reduces MAE by roughly 1.5 years compared to MT‐AE and by 1.2 years versus DS‐AE.

\begin{figure*}[!p]  
  \centering

  \begin{subfigure}[t]{0.48\textwidth}
    \centering
    \includegraphics[width=\linewidth,height=0.4\textheight,keepaspectratio]{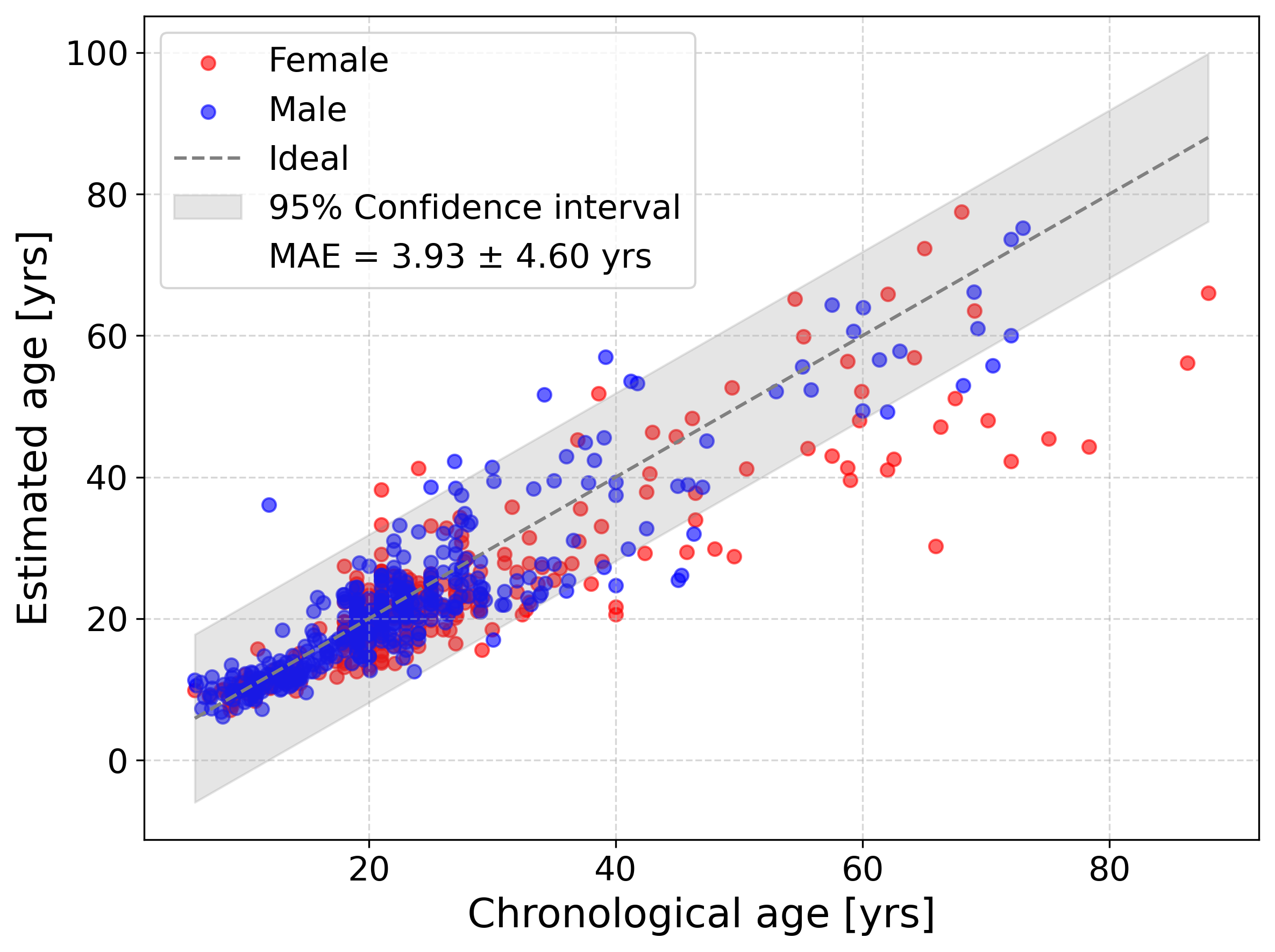}
    \caption{Baseline (3D ResNet)}
  \end{subfigure}\hfill
  \begin{subfigure}[t]{0.48\textwidth}
    \centering
    \includegraphics[width=\linewidth,height=0.4\textheight,keepaspectratio]{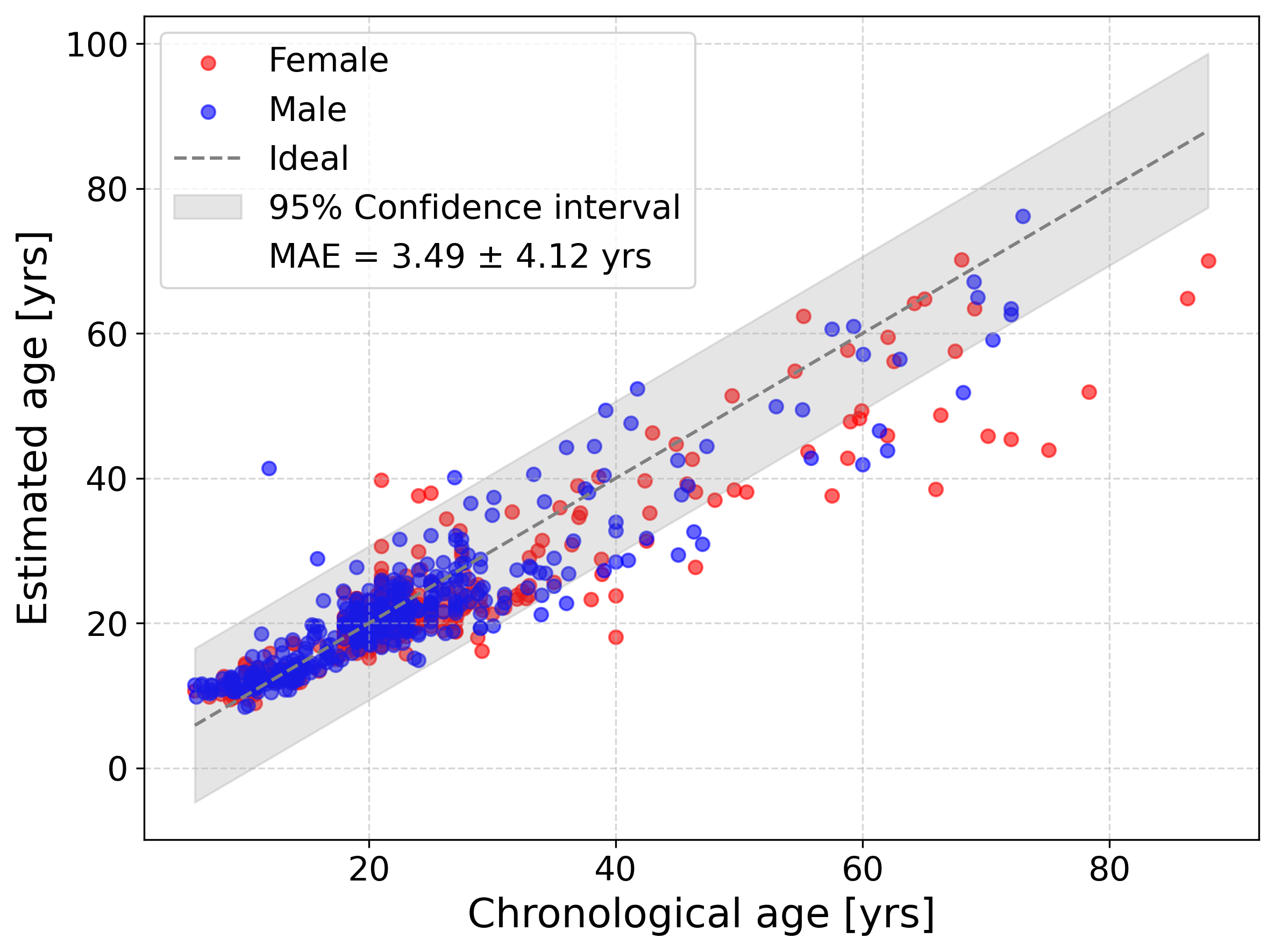}
    \caption{Autoencoder}
  \end{subfigure}

  \medskip
  \begin{subfigure}[t]{0.48\textwidth}
    \centering
    \includegraphics[width=\linewidth,height=0.4\textheight,keepaspectratio]{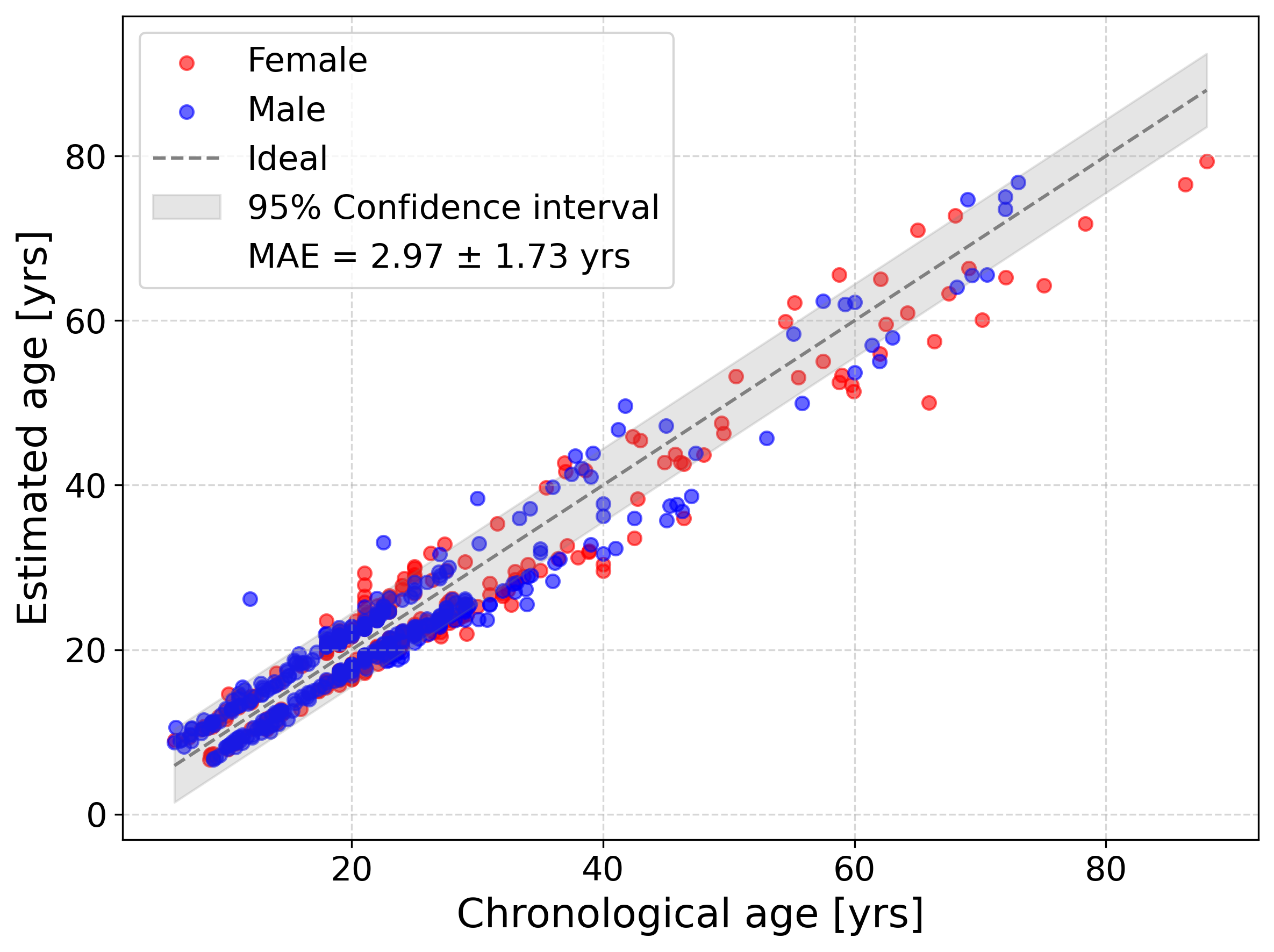}
    \caption{Multitask Autoencoder}
  \end{subfigure}\hfill
  \begin{subfigure}[t]{0.48\textwidth}
    \centering
    \includegraphics[width=\linewidth,height=0.4\textheight,keepaspectratio]{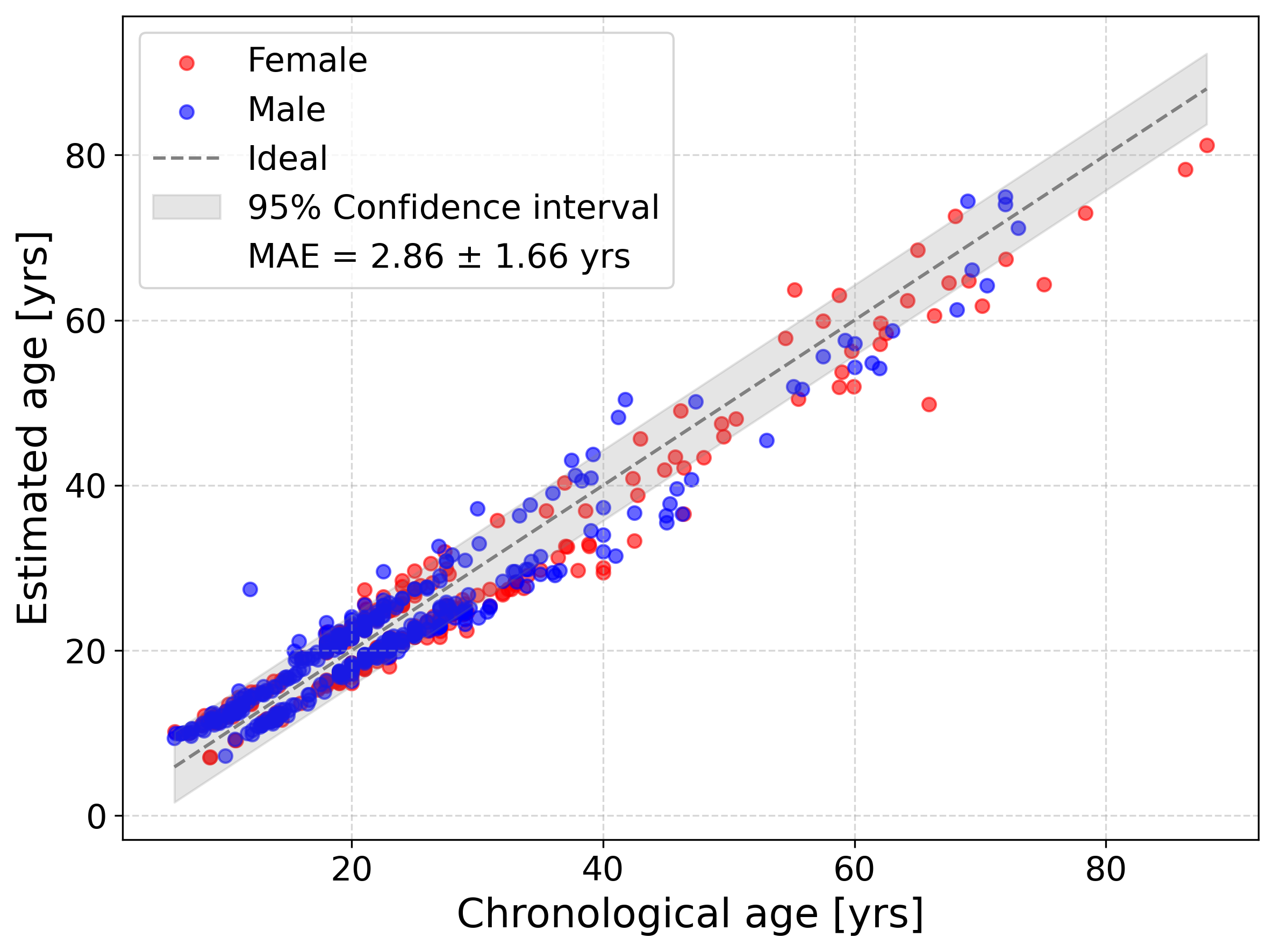}
    \caption{{Deeply Supervised Autoencoder}}
  \end{subfigure}

  \medskip
  \makebox[\textwidth][c]{%
    \begin{subfigure}[t]{0.55\textwidth}
      \centering
      \includegraphics[width=\linewidth,height=0.35\textheight,keepaspectratio]{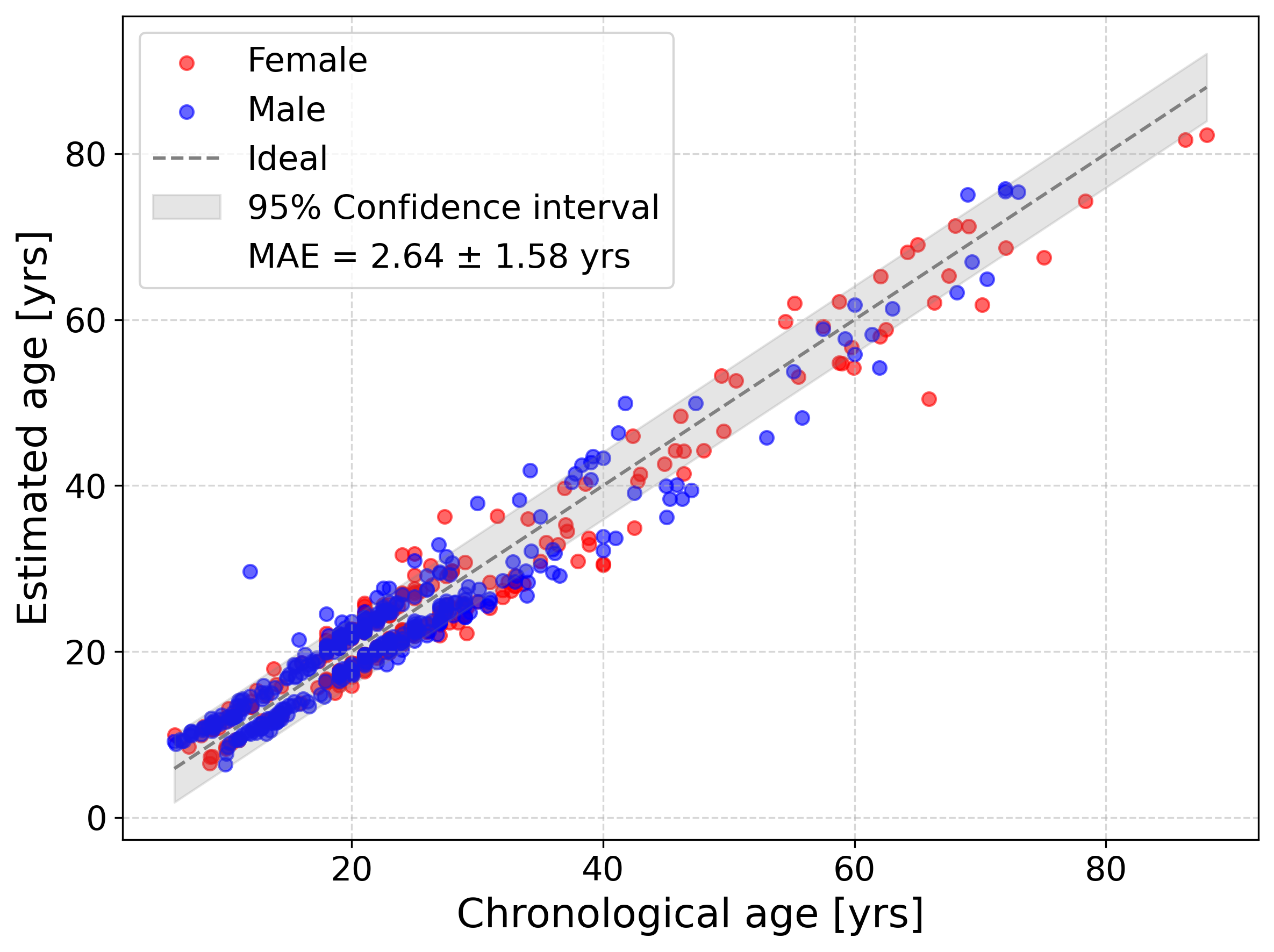}
      \caption{Our proposed DSMT-Autoencoder}
    \end{subfigure}%
  }

  \caption{Comparison of predictions for different models: (a) Baseline (3D ResNet), (b) Autoencoder, (c) Multitask autoencoder, (d) Deeply supervised autoencoder, and (e) our proposed Deeply Supervised Multitask autoencoder. Each graph plots predicted brain age versus chronological brain age, with sex and confidence intervals indicated.}
  \label{models_comparison}
\end{figure*}

Notably, the error bars for DSMT‐AE remain narrower at every age, indicating more consistent performance even in sparse or high‐variance strata. In the youngest group ($\le 25$), all three models perform comparably (MAE $\approx 2.3\,$yrs), but DSMT‐AE still exhibits the tightest dispersion (SD $\approx 0.5\,$yrs).  Through the middle‐age ranges (36–55\,yrs), DSMT‐AE maintains a 10–15\% MAE advantage over MT‐AE and reduces variance by 20–25\%, underscoring how deep supervision and sex‐aware signals improve feature discrimination across the adult lifespan.  For the oldest groups (66–75 and $>75$\,yrs), where sMRI patterns grow more heterogeneous, DSMT‐AE’s multitask regularization proves especially valuable: it cuts MAE by up to 17\% and yields the smallest confidence intervals of all variants.

These age‐stratified results confirm that DSMT‐AE not only delivers superior average accuracy but also sustains robustness across wide demographic ranges.  By enforcing both sex classification and hierarchical supervision, our framework learns latent representations that generalize consistently, even in age brackets with higher biological variability.

\section{Conclusions and Future Works}
\label{con}
In this work, we have introduced a novel \emph{Deeply Supervised Multitask Autoencoder} (DSMT‐AE) for brain age estimation from volumetric T1‐weighted MRI.  By jointly optimizing three complementary objectives, i.e., image reconstruction, sex classification, and age regression—across multiple encoder depths, DSMT‐AE learns rich, anatomically informed features while avoiding vanishing‐gradient issues.  Our extensive experiments on the publicly available OpenBHB cohort demonstrate that DSMT‐AE not only sets a new state-of-the-art in mean absolute error and explained variance but also yields robust, consistent predictions across both male and female subgroups.  Ablation studies confirm that each component, i.e., unsupervised reconstruction, demographic regularization, and deep supervision, contributes uniquely to the final performance, and that their combination produces a synergistic improvement over any single technique. Future work includes validating DSMT-AE on cohorts with mild cognitive impairment and neurodegenerative diseases, such as Alzheimer’s and Parkinson’s, to assess its sensitivity to disease-related structural changes.


\section{Declaration of competing interest}
The authors declare that they have no known competing financial interests or personal relationships that could have appeared to influence the work reported in this paper.

\section{Code and Data availability}
The source code developed in this study is made publicly available. The data utilized for experimentation is the OpenBHB dataset, a publicly accessible multi-site brain MRI dataset for age prediction and debiasing. It can be directly downloaded from the official IEEE DataPort repository. 

\section{Acknowledgments}
This research was supported by the MSIT(Ministry of Science and ICT), Korea, under the ITRC (Information Technology Research Center) support program (IITP-2025-RS-2020-II201789), and the Artificial Intelligence Convergence Innovation Human Resources Development (IITP-2025-RS-2023-00254592) supervised by the IITP (Institute for Information \& Communications Technology Planning \& Evaluation).



\end{document}